\documentclass[runningheads]{llncs}
\usepackage{graphicx}
\usepackage{amsmath}
\usepackage{amssymb}
\usepackage{bm}

\newcommand{\norm}[1]{\lVert#1\rVert}

\begin{document}
\title{GANDALF: Generative Adversarial Networks with Discriminator-Adaptive Loss Fine-tuning for Alzheimer's Disease Diagnosis from MRI}
\titlerunning{GANDALF for MRI-based Alzheimer's Disease Diagnosis}
\author{Hoo-Chang Shin\inst{1}, Alvin Ihsani\inst{1}, Ziyue Xu\inst{1}, Swetha Mandava\inst{1}, Sharath Turuvekere Sreenivas\inst{1}, Christopher Forster\inst{1}, 
Jiook Cha\inst{2}\\ \and Alzheimer's Disease Neuroimaging Initiative} %\and

% index{Shin, Hoo-Chang}
% index{Xu, Ziyue}
% index{Ihsani, Alvin}
% index{Mandava, Swetha}
% index{Sreenivas, Sharath Turuvekere}
% index{Forster, Christopher}
% index{Cha, Jiook}

\authorrunning{Shin et al.}
\institute{NVIDIA Corporation;
\email{hshin@nvidia.com}\and
Department of Psychology, Center for REAL Intelligence, AI Institute, Seoul National University;
\email{connectome@snu.ac.kr}}
% \institute{***********}
% \institute{Nvidia Corporation}
%
\maketitle              % typeset the header of the contribution
\begin{abstract}

Positron Emission Tomography (PET) is now regarded as the gold standard for the diagnosis of Alzheimer's Disease (AD).
However, PET imaging can be prohibitive in terms of cost and planning, and is also among the imaging techniques with the highest dosage of radiation.
%: the short half-life of the radioisotopes requires on-site production in remote areas, no-show patients result in radioisotopes being wasted, the length of imaging sessions is determined by the tracer being used, and small variations ($\sim$5 min) in the acquisition start time may cause over- or under-estimation of quantitative parameters.
Magnetic Resonance Imaging (MRI), in contrast, is more widely available and provides more flexibility when setting the desired image resolution.
%and determining the length of the acquisition so as to minimize artifacts (e.g. motion).

Unfortunately, the diagnosis of AD using MRI is difficult due to the very subtle physiological differences between healthy and AD subjects visible on MRI. As a result, many attempts have been made to synthesize PET images from MR images using generative adversarial networks (GANs) in the interest of enabling the diagnosis of AD from MR.
Existing work on PET synthesis from MRI has largely focused on Conditional GANs, where MR images are used to generate PET images and subsequently used for AD diagnosis. There is no end-to-end training goal.

This paper proposes an alternative approach to the aforementioned, where AD diagnosis is incorporated in the GAN training objective to achieve the best AD classification performance. Different GAN losses are fine-tuned based on the discriminator performance, and the overall training is stabilized. The proposed network architecture and training regime show state-of-the-art performance for three- and four- class AD classification tasks.

\end{abstract}

%%%%%%%%%%%%%%%%%%%%%%%%%%%%%%%%%%%%%%%%%%%%%%%%%%%%%%%%
\section{Introduction}

\subsection{PET Imaging and AD Diagnosis}

\begin{figure}[h]
    \centering
    \includegraphics[width=1\textwidth]{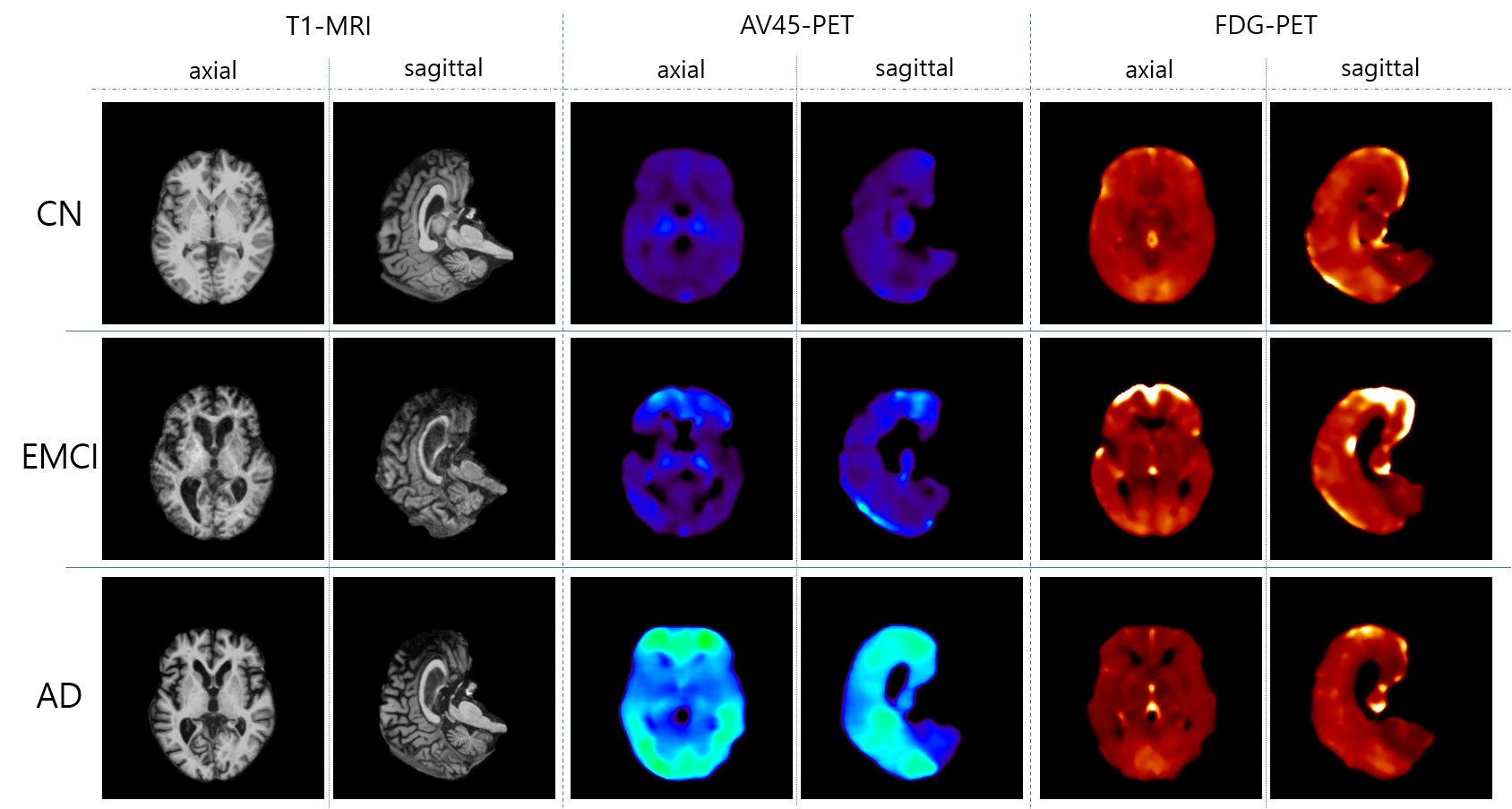}
    \caption{
    Examples of T1-weighted MRI, AV45-PET, FDG-PET brain images of Cognitive Normal (CN), early mild cognitive impairment (EMCI), and Alzheimer's disease (AD).
    The differences are much more clearly visible on the PET images than in the MR images, especially for the EMCI case.
    On the EMCI case, increased accumulation of AV45 in the medial temporal, occipital, and frontal lobe (inferior frontal gyrus shown) is noticeable.
    AD shows reduced brain metabolism in the FDG scan, and significant uptake of AV45 compared to both CN and EMCI due to widespread accumulation of Amyloid-$\beta$.
    In the T1-weighted MRI images, the size increment of the ventricles is visible from CN to EMCI to AD, however it is not as clearly visible as in PET.
    % CN shows high uptake of FDG signifying high-brain metabolism, and no significant accumulation of AV45; EMCI shows similar brain metabolism to CN based on FDG, but increased accumulation of AV45 in the medial temporal, occipital, and frontal lobe (inferior frontal gyrus shown) signifying accumulation of Amyloid-$\beta$ at these sites; AD shows reduced brain metabolism in the FDG scan, and significant uptake of AV45 compared to both NC and EMCI due to widespread accumulation of Amyloid-$\beta$. T1-weighted MRI images also show differences between the patients, where increased brain atrophy is visible from NC to EMCI to AD; this is especially visible in the increase of the size of ventricles.
    }
    \label{fig:ad_cn_pet_mri_examples}
\end{figure}

Alzheimer's Disease (AD) is the most common cause of dementia, affecting quality of life for many elderly people and their families. Early diagnosis and intervention of AD can improve the quality of life by significantly slowing the progression of the disease, thus it is an active area of research. 
Positron Emission Tomography (PET) appears to be a very promising imaging technique to assess the progression and stage of the disease by monitoring the spread of Tau-protein in the form of Neurofibrillary Tangles (NFT) and Amyloid beta (A$\beta$)~\cite{Johnson16,Johnson13,Schwartz16}. As a functional imaging technique, PET uses a radioactive tracer injected into the patient, and images the distribution of the tracer over the course of minutes or hours.
% The tracer is designed with a preferential biding for a ligand indicative of the pathology of interest.
% Positron Emission Tomography (PET) is an imaging technique where a radioactive tracer is injected to the body and its activity is imaged over time.
% This allows in-vivo assessment of brain function and pathology changes and a disease can be sometimes detected before it can be visualized on other imaging tests.
% PET is also showing great potential in supporting early clinical diagnosis of dementia and its early intervention.
% Specifically to AD, the tracer AV-1451 (T-807) was developed to bind to Tau protein, however, it binds longer NFT which have been implicated in the onset of AD~\cite{Johnson16}.
% The AV-1451 binding affinity to NFT enables the detection of NFT buildup, and the analysis of the relationship between NFT accumulation to functional and anatomical changes in the brain, and the eventual appearance of symptoms typical of AD.
In AD research, PET imaging techniques measure amyloid plaque (AV45)~\cite{berti2011pet,sevigny2016antibody} and tau protein aggregates (AV1451)~\cite{marquie2015validating,xia201318f} that are essential to understanding AD pathology and diagnosis.
Compared to AV45-/AV1451- PET, FDG-PET usually helps differentiate AD from other causes of dementia, because it can characterize the patterns of glucose metabolism in the brain that are specific to AD~\cite{marcus2014brain}.
Example T1-weighted Magnetic Resonance images, AV45-/FDG- PET brain images of CN and AD are shown in Figure~\ref{fig:ad_cn_pet_mri_examples}.

While PET plays an important role for AD diagnosis, it can be prohibitive in terms of cost and planning: \textit{(1)} the short half life of the radioisotopes requires on-site production in remote regions; \textit{(2)} no-show patients result in radioisotopes being wasted; \textit{(3)} the length of imaging sessions is determined by the tracer and the use case so motion artifacts may be unavoidable, and lastly; \textit{(4)} small variations ($\sim$5 min) in the acquisition start time may cause over- or under-estimation of quantitative parameters.
It is also not as widely available as Magnetic Resonance Imaging (MRI).

\subsection{Synthesizing PET Images from MR for AD Diagnosis}

To address the shortcomings of PET for AD diagnosis, a number of studies have attempted AD diagnosis from T1-weighted MR images.
While T1-weighted MRI is most suitable for visualizing anatomical structures in the brain, it is not optimal for AD diagnosis because it does not highlight functional or metabolic properties of brain tissues.
The question arises as to whether one can leverage existing combined PET-MR image pairs (a combined imaging modality available to only large research institutions) to generate PET images from MR-only image acquisitions.

Conditional generative adversarial networks (CGAN)~\cite{isola2017image} have previously been used to generate images of a modality from a paired input image of a different modality.
Frequent examples of such paired images are images and label maps, images and sketch, and pictures of the same scene from one lighting condition to another (e.g. day/night).
For medical image analysis, such as AD diagnosis, PET image is generated from MRI using CGAN.
The generated PET is then used to train AD classification network.

This work proposes an approach similar to CGAN, where CGAN is trained end-to-end with the final goal of AD classification.
If trained with classification goal, then the performance of generating realistic images may be compromised.
We overcome this limitation by adaptively fine-tuning the GAN losses and classification losses.
Also, the overall GAN training is stabilized by the loss fine-tuning.
State-of-the-art result on three- and four- class AD classification are achieved with the proposed architecture and training regime.

\subsection{Dataset and Classification of Cognitive Decline}

We use the publicly available ADNI (Alzheimer’s Disease Neuroimaging Initiative) dataset comprised of F18-AV-45 (florbetapir) and F18-FDG (fluorodeoxyglucose) PET image pairs along with the co-registered T1-weighted MRI.
The dataset contains six dementia related conditions: cognitive normal (CN), early mild cognitive impairment (EMCI), late mild cognitive impairment (LMCI), mild cognitive impairment (MCI), subjective memory complaint (SMC), and Alzheimer disease (AD).
Among these conditions, SMC is difficult to subjectively distinguish from CN.
Also, there may be overlaps between EMCI/LMCI and MCI.
Therefore, we test binary classification of AD/CN, three- and four- class classification of AD/MCI/CN and AD/LMCI/EMCI/CN for early AD diagnosis.
Figure~\ref{fig:ad_cn_pet_mri_examples} show some examples of CN, EMCI, AD images in the dataset.

We randomly divide the dataset with 70\% training, 10\% validation, and 20\% testing according to the patients, resulting 722/104/207 subjects for each train/validation/test set.
Some subjects have multiple scans (i.e., more than one temporal scan), with the total 1,525 image triplets (AD45-/FDG-PET, T1-MRI).
% The number of image triplet for each condition are: CN\,428, EMCI\,456, LMCI\,219, MCI\,158, SMC\,102, AD\,162.
The images are pre-processed using FreeSurfer~\cite{FreeSurfer}.
The T1-weighted images are skull-stripped, where non-cerebral matters such as skull and scalp are removed.
Registration, re-scaling~\cite{greve2014cortical} and partial volume correction~\cite{greve2016different} is applied to the PET images.
The T1-weigthed images are re-scaled to $1\textrm{mm}^{3}$ with $256^{3}$ voxels, and PET images are $2\times 93\times 76\times 76$ voxels with 2 temporal resolution.
% at $T\textrm{mm}\times 1\textrm{mm}\times 1\textrm{mm}$ resolution for axial/sagittal/coronal views respectively.
% Here, $T$ is the length of time-points in PET imaging.
% They are mostly 2-4 in our dataset, where some images have single time point or more than four time points.
% We only select $T=2$ and for images with $T>2$ we use the first two time points.
% For training, each image is normalized per-instance, by subtracting its mean and dividing its three times standard deviation.

%%%%%%%%%%%%%%%%%%%%%%%%%%%%%%%%%%%%%%%%%%%%%%%%%%%%%%%%
\section{Related Works}

Image-based AD diagnosis is regarded as a challenging task.
Most of the prior works use a combination of structural and functional imaging, such as T1-weighted MRI and PET, or T1-weighted MRI and functional MRI such as DTI (diffusion tensor imaging).
They also typically focus on binary classification of each state category, such as AD vs. NC, or AD vs. MCI.

A combination of T1-weighted MRI, AV45-/FDG-PET was used with multi-feature kernel supervised within-class-similar discriminative dictionary learning algorithm to demonstrate binary classification of AD/NC, MCI/NC, AD/MCI in \cite{li2018classification}.
A combination of T1-weighted MRI and FDG-PET with three-dimensional convolutional neural network (CNN) was used to demonstrate binary classification of CN/AD, CN/pMCI, sMCI/pMCI in \cite{huang2019diagnosis}.
GAN was used to generate additional PET images from T1-weighted MRI that do not have AV45-PET image pairs in \cite{yan2018generation}.
MRI and real-/synthetic- PET image pairs are subsequently used to train CNN to perform binary classification of stable-MCI/progressive-MCI.

Functional MRI (fMRI) is an MRI imaging technique most similar to PET that it can measure brain activity by detecting changes associated with blood flow.
A minimum spanning tree (MST) classification framework was proposed in \cite{cui2018classification} to perform binary classification of MCI/NC, AD/CN, and AD/MCI using fMRI.
A combination of T1-weighted MRI and Diffusion Tensor Imaging (DTI) was used with Multiple Kernel Learning to demonstrate binary classification of CN/AD, CN/MCI, AD/MCI in \cite{ahmed2017recognition}.

More recent work demonstrates diagnosing AD from T1-weighted MRI only. 
Longitudinal studies with landmark-based features and support vector machines to classify CN/AD and CN/MCI in \cite{zhang2017alzheimer}.
T1-weighted MRI was used with convolutional autoencoder based unsupervised learning for the CN/AD and progressive-MCI/stable-MCI classification task in \cite{oh2019classification}.
Other recent works show multi-class classification using T1-weighted MRI. A variant of DenseNet CNN was used for multi-class classification of AD/MCI/NC using MRI in \cite{wu2019learning}.
T1-weighted MRI was used with CNN to demonstrate binary classification of NC/AD and three-class classification of NC/AD/MCI in \cite{esmaeilzadeh2018end}.

%%%%%%%%%%%%%%%%%%%%%%%%%%%%%%%%%%%%%%%%%%%%%%%%%%%%%%%%
% \section{Dataset and Pre-Processing}

% We use the ADNI dataset~\cite{jack2008alzheimer} that is widely used for imaging-based AD studies.
% The initial phase of the ADNI data contained 800 participants with late MCI.
% It was followed by a second phase of ADNI (ADNI-2) to include patients with early MCI.
% From the entire ADNI dataset we selected subjects with three imaging modalities, namely T1-weighted MRI, AV45-PET, and FDG-PET, resulting in 1,033 unique individuals with six dementia conditions: CN, EMCI, LMCI, MCI, SMC, AD.

%%%%%%%%%%%%%%%%%%%%%%%%%%%%%%%%%%%%%%%%%%%%%%%%%%%%%%%%
\section{Methods}

The \texttt{pix2pix}~\cite{isola2017image} CGAN architecture is widely adopted in the medical image analysis domain for synthesizing from one image modality to another.
For instance, Yan et. al~\cite{yan2018generation} use the CGAN to generate AV45-PET from T1-weighted MRI to supplement the training dataset with additional synthetic PET-MRI image pairs.
While for generating an image of different modality may be an end-goal for computer vision domain, in medical domain we often want to diagnose a disease, such as AD, using the generated image.
We hypothesize that a GAN designed and trained with this diagnosis end-goal in mind can outperform in AD diagnosis, compared to other types of CGAN application where synthesis and diagnosis are trained separately.

\subsection{Conditional Generative Adversarial Networks}

The \texttt{pix2pix}~\cite{isola2017image} CGAN is trained with the following objective:

\begin{align}
    G^*  = \arg\min_G\max_D \mathcal{L}_{cGAN}(G,D) + \lambda \mathcal{L}_{L1}(G).\label{full_objective}
\end{align}
where $\mathcal{L}_{cGAN}(G,D)$ and and $\mathcal{L}_{L1}(G)$ are defined as

\begin{align}
    \mathcal{L}_{cGAN}(G,D) =& \mathbb{E}_{x,y}[\log D(x,y)] + 
                 \mathbb{E}_{x,z}[\log (1-D(x,G(x,z))],\\
    \mathcal{L}_{L1}(G) =& \mathbb{E}_{x,y,z}[\norm{y-G(x,z)}_1].
    \label{cGAN_L1_equation}
\end{align}

\noindent
where $x$, $y$ and $G(x,z)$ can be regarded as MRI, PET input, and generated PET.
The CGAN consists of a generator ($G$) that has encoder-decoder architecture, and a discriminator ($D$) that is a CNN classifier.
The U-Net~\cite{ronneberger2015u} architecture is usually used as the $G$ that takes an input image and generates an output image of a same size but of different modality or characteristics.
PET conventionally has lower image resolution than MRI, so we modify the U-Net architecture to take the different resolutions into account - MRI: $256\times 256\times 256$; PET: $2\times 93\times 76\times 76$.
The encoder part has eight layers while the decoder part has five.
Only the middle five layers in the encoder-decoder part has the skip-connection, with the last two up-sampling (transpose convolution) layers to make the target PET resolution.
% The discriminator is a CNN that takes MRI and PET as input.
The discriminator CNN has three convolutional (conv-) layers that take MRI input, and two conv-layers that take PET input.
The two branches of conv-layers are merged and followed by two additional conv-layers for classification.

\begin{figure}[h]
    \centering
    \includegraphics[width=1\textwidth,trim=2.6cm 13.5cm 1cm 3.9cm,clip]{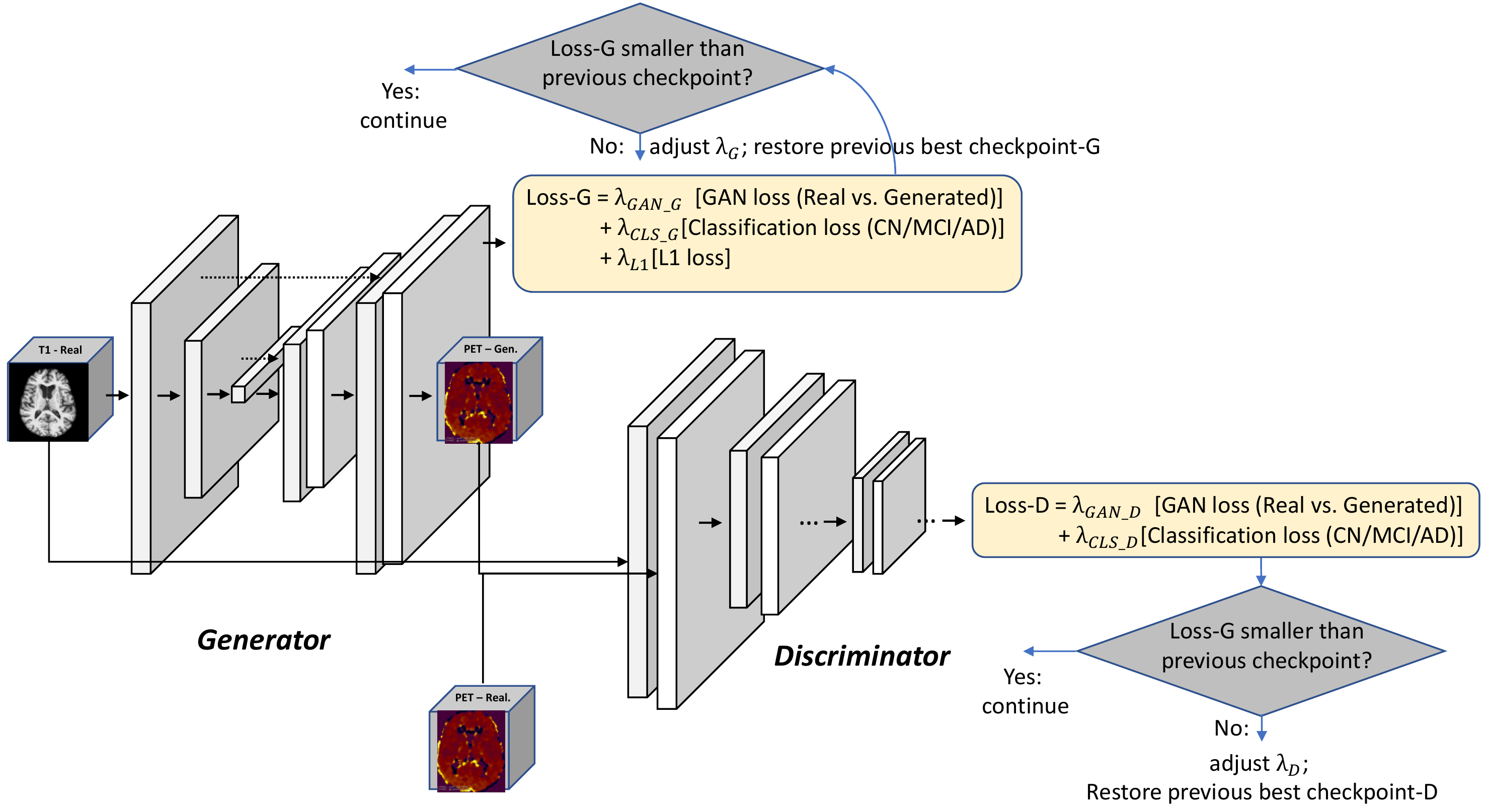}
    \caption{Overall architecture and training pipeline. While generator and discriminator are trained independently to compete against each other, they are both trained with the additional AD classification loss that are adjusted \textit{(1)} to generate realistic PET images, and \textit{(2)} to perform well on AD classification.
    In addition, losses are monitored and weights are adjusted to stablilize the GAN training, preventing loss oscillation.}
    \label{fig:gandalf_arch}
\end{figure}

\subsection{GAN with Discriminator-Adaptive Loss Fine-tuning}

GAN is trained with minimax objective~\cite{goodfellow2014generative} where $G$ and $D$ compete with each other.
CGAN is trained with an additional $L1$ loss for the $G$, and a patch-GAN~\cite{isola2017image} classifier for the $D$.
%that penalizes structure at the scale of smaller image patches.
% While they are sufficient for generating synthetic images that are realistic, in medical image analysis we usually have an additional end-goal, that is to diagnose disease.
% We hypothesize that training the GAN with the additional AD classification objective will result in better generative model for AD diagnosis - also when we synthesize a different image modality from an input image for better diagnosis.
% Our generative network is trained the same way as CGAN, while our 
The $D$ in our generative network is trained with additional AD classification losses: \textit{(1)} based on real MRI and generated PET input, multiplied by a hyper-parameter $\lambda_{\textrm{GAN}_D}$, and \textit{(2)} based on real MRI and PET, multiplied by $\lambda_{\textrm{CLS}_D}$:

\begin{align}
    \mathcal{L}_\textrm{D}(D,G) = &\lambda_{\textrm{GAN}_D}\arg\min_G \mathcal{L}_{cGAN}(G,D) + 
    \lambda_{\textrm{CLS}_D}\mathbb{E}_{x,y,\hat{y}}[\log D(\hat{y}|x,y)]\label{GANDALF_equation_D},
\end{align}

\noindent
where $\hat{y}$ is the AD label.

The $G$ is also trained with AD classification loss based on real MRI and generated PET input, in addition to the GAN loss and $L1$ loss.
Each loss is multiplied with hyper-parameters to control their relative importance during the training - $\lambda_{\textrm{CLS}_G}$, $\lambda_{\textrm{GAN}_G}$, and $\lambda_{L1}$:

\begin{align}
    \mathcal{L}_\textrm{G}(G,D) = 
    &\lambda_{\textrm{GAN}_G}\arg\max_D \mathcal{L}_{cGAN}(G,D) + \nonumber \\
    &\lambda_{\textrm{CLS}_G}\mathbb{E}_{x,\hat{y}}[\log D(\hat{y}|x,G(x,z))] + 
    \lambda_{L1}\mathcal{L}_{L1}(G).
                 \label{GANDALF_equation_G}
\end{align}

In the earlier phase of the GAN training, generated PET likely are far from the real ones.
They progressively become more realistic as the training proceeds.
Therefore, $D$ is trained initially with small $\lambda_{\textrm{GAN}_D}$ and gradually increased during the training, while $\lambda_{\textrm{CLS}_D}$ starts from a larger value and gradually decreased.
This encourages the $D$ to focus on AD classification when $G$ is improving to generate more realistic PET images.
The $G$ is trained with a large $\lambda_{\textrm{GAN}_G}$ at first so it can focus on generating realistic PET in the beginning.
It is gradually decreased as $\lambda_{\textrm{CLS}_G}$ increases from a smaller value, to emphasize AD classification using the generated PET images.
We set  $\lambda_{\textrm{GAN}_D}$ and $\lambda_{\textrm{CLS}_G}$ as 0.01 and linearly increase 10 times per epoch, $\lambda_{\textrm{CLS}_D}$ and $\lambda_{\textrm{GAN}_G}$ initially as 100 and decrease 1/10 times per epoch.
We train for 1000 epochs using ADAM optimizer~\cite{kingma2014adam}.

\subsubsection{Stabilizing Training}

Training $D$ and $G$ independently, the $D$ and $G$ loss can oscillate rather than being in a stable convergence state~\cite{ArjovskyB17}.
To remedy this problem we continuously monitor the $D$ and $G$ loss, and adjust the hyper-parameters $\lambda$ for the losses if any one is lower compared to the previous epoch.
Loss oscillation generally occurs when the training has well proceeded, and this is when AD classification losses get higher weights.
This is similar to the approach of \cite{roth2017stabilizing} penalizing $D$ weights with annealing to stabilize the GAN training.
For example, when the $D$ loss starts to oscillate and becomes higher compared to the previous epoch, then \textit{(1)} its previous checkpoint is restored, and \textit{(2)} $\lambda_{\textrm{CLS}_D}$ gets decreased.
The overall training pipeline is shown in Figure~\ref{fig:gandalf_arch}.

%%%%%%%%%%%%%%%%%%%%%%%%%%%%%%%%%%%%%%%%%%%%%%%%%%%%%%%%
\section{Results}

We perform two- to four- class AD classification using T1-weighted MRI input.
The two-class AD classification results is shown in Table~\ref{tab:ADvsCN_comparison}.
The CNN approach in \cite{esmaeilzadeh2018end} report better performance on the two-class AD/CN classification, and \texttt{GANDALF} show similar performance to \texttt{pix2pix + CNN} method.
We suspect this may be because AD vs. CN is more clearly visible than AD/MCI/CN or AD/LMCI/EMCI/CN on MRI, so a deep CNN with good hyper-parameter set can provide better result and PET plays a rather limited role for the diagnosis.
We did not conduct a thorough hyper-parameter search for \texttt{GANDALF} in this study.

\begin{table}
\caption{Comparison of MRI-based AD diagnosis for AD vs. CN binary classification. CNN based method~\cite{esmaeilzadeh2018end} reports best performance which may indicate using PET and synthesized PET is more useful for early AD diagnosis.}\label{tab:ADvsCN_comparison}
\begin{center}
\begin{tabular}{|l||c||c|c|c|c|}
\hline
Method & AD/CN & Acc & F\textsubscript{2} & Prec & Rec \\
\hline
Esmaeilzadeh et.al~\cite{esmaeilzadeh2018end} &  200/230 & $\bm{94.1}$ & $\bm{0.93}$ & $\bm{0.92}$ & $\bm{0.94}$  \\
\hline
\texttt{pix2pix} + \texttt{CNN} & 162/428 & 85.2 & 0.83 & 0.84 & 0.83   \\
\hline
\texttt{GANDALF} & 162/428 & 85.2 & 0.84 & 0.84 & 0.84 \\
\hline
\end{tabular}
\end{center}
\end{table}

Results of the three-class AD/MCI/CN classification task are shown in Table~\ref{tab:three-class_comparison}.
We achieve state-of-the-art performance on the three-class classification compared to the prior works using T1-weighted MRI input.
MCI may show more subtle difference on the MRI compared to the AD as can be seen in Figure~\ref{fig:ad_cn_pet_mri_examples}.
This, and the consistent better performance of the generative methods compared to the prior works could indicate that an additional training of synthesizing PET can help achieving better performance for early AD diagnosis.

\begin{table}
\caption{MRI-based AD diagnosis for AD/MCI/CN three-class classification. Better performance shown by generative methods may suggest additional training to generate synthesized PET can be promising for early diagnosis of AD using MRI.}\label{tab:three-class_comparison}
\begin{center}
\begin{tabular}{|l||c||c|c|c|c|}
\hline
Method & AD/MCI/CN & Acc & F\textsubscript{2} & Prec & Rec \\
\hline
Esmaeilzadeh et.al~\cite{esmaeilzadeh2018end} & 200/411/230 & 61.1 & 0.62 & 0.59 & 0.63 \\
\hline
Wu et.al~\cite{wu2019learning} & 130/455/200 & N/A & 0.49 & 0.62 & 0.35 \\
\hline
\texttt{pix2pix} + \texttt{CNN} & 162/456/428 & 71.3 & 0.63 & 0.64 & 0.63 \\
\hline
\texttt{GANDALF} & 162/456/428 & $\bm{78.7}$ & $\bm{0.69}$ & $\bm{0.83}$ & $\bm{0.66}$ \\
\hline
\end{tabular}
\end{center}
\end{table}

Lastly, four-class classification of AD/LMCI/EMCI/CN results are shown in Table~\ref{tab:four-class_comparison}.
We show a meaningful first result on classifying early-MCI and late-MCI from CN and AD, a promising first step for early AD diagnosis using T1-weighted MRI.
Our proposed \texttt{GANDALF} method also shows improved performance compared to the \texttt{pix2pix + CNN} method.
Towards the end of the \texttt{GANDALF} training, the entire network acts as a classification network with T1-weighted MRI input.
Finding a better/deeper classifier/discriminator architecture could improve the final classification performance.
However this should be balanced with the generator architecture/depth for the GAN training with the minimax objective.
A thorough hyper-parameter search could also improve the final performance.

\begin{table}
\caption{MRI-based AD diagnosis for AD/LMCI/EMCI/CN four-class classification. We show meaningful result that can be promising for early diagnosis on AD using T1-weighted MRI input.}\label{tab:four-class_comparison}
\begin{center}
\begin{tabular}{|l||c||c|c|c|c|}
\hline
Method & AD/LMCI/EMCI/CN & Acc & F\textsubscript{2} & Prec & Rec \\
\hline
\texttt{pix2pix} + \texttt{CNN} & 162/456/219/428 & 33.0 & 0.34 & 0.34 & 0.34 \\
\hline
\texttt{GANDALF} & 162/456/219/428 & $\bm{37.0}$ & $\bm{0.40}$ & $\bm{0.39}$ & $\bm{0.40}$ \\
\hline
\end{tabular}
\end{center}
\end{table}

%%%%%%%%%%%%%%%%%%%%%%%%%%%%%%%%%%%%%%%%%%%%%%%%%%%%%%%%
\section{Conclusion}

Early diagnosis and intervention of Alzheimer's disease (AD) can significantly slow the progression of the disease and improve patients' condition and the life quality of the patient and their caregivers.
PET imaging can provide great insight for early diagnosis of AD, however, it is rarely available outside of research environments.
Earlier works on MRI-based AD diagnosis use conditional generative adversarial networks (GAN) to synthesize PET from MRI, and subsequently use the generated PET for AD diagnosis.

We propose a network where AD diagnosis end-goal is incorporated into the MRI-PET synthesis and trained end-to-end, instead of first synthesizing PET and then use it for AD diagnosis.
Furthermore, we suggest a training scheme to stabilize the GAN training.
We achieve state-of-the-art MRI-based AD diagnosis for three-class AD classification of AD/MCI/CN.
We also achieve the first meaningful result on four-class (AD/LMCI/EMCI/CN) classification that can be promising for early diagnosis of AD based on MRI, to the best of our knowledge.

%%%%%%%%%%%%%%%%%%%%%%%%%%%%%%%%%%%%%%%%%%%%%%%%%%%%%%%%

\section*{Acknowledgement}
The authors would like to thank Seonjoo Lee of Columbia University Medical Center for the discussion and help in data pre-processing.

%%%%%%%%%%%%%%%%%%%%%%%%%%%%%%%%%%%%%%%%%%%%%%%%%%%%%%%%

\bibliographystyle{splncs04}
\bibliography{1826}

\end{document}